\documentclass[aps,twocolumn,superscriptaddress]{revtex4}
\usepackage{graphicx}
\usepackage{amsmath}
\usepackage{amsfonts}
\usepackage{amssymb}
\usepackage{psfrag}
\usepackage{wrapfig}
\usepackage{subfigure}
\usepackage{makeidx}
\usepackage{bm}
\usepackage{epsf}
\usepackage{multirow}
\usepackage{upgreek}
\usepackage{ulem}
\usepackage[colorlinks,urlcolor=blue,citecolor=blue]{hyperref}
\DeclareMathOperator{\sech}{sech}

\begin{document}
\title{Self-Adapted Josephson Oscillation of Dark-Bright Solitons under Constant Forces}
\author{Ling-Zheng Meng}
\thanks{These authors contribute equally.}
\affiliation{School of Physics, Northwest University, Xi'an, 710127, China}
\affiliation{School of Science, Xi'an Technological University, Xi'an 710032, China}
\author{Xi-Wang Luo}
\thanks{These authors contribute equally.}
\affiliation{CAS Key Laboratory of Quantum Information, University of Science and Technology of China, Hefei, Anhui 230026, China}
\affiliation{Synergetic Innovation Center of Quantum Information and Quantum Physics, University of Science and Technology of China, Hefei, Anhui 230026, China}
\affiliation{Hefei National Laboratory, University of Science and Technology of China, Hefei 230088, China}
\author{Li-Chen Zhao}
\email{zhaolichen3@nwu.edu.cn}
\affiliation{School of Physics, Northwest University, Xi'an, 710127, China}
\affiliation{NSFC-SPTP Peng Huanwu Center for Fundamental Theory, Xi'an 710127, China}
\affiliation{Shaanxi Key Laboratory for Theoretical Physics Frontiers, Xi'an 710127, China}
\date{\today}
\begin{abstract}
We study the propagation of dark-bright solitons in two-component Bose-Einstein condensates (BECs) with general nonlinear parameters, and explore how nonlinear interactions enrich the soliton dynamics giving rise to nonsinusoidal oscillations under constant forces. Treating the bright soliton as an effective barrier, we reveal that such oscillations are characterized by the Josephson equations with self-adapted critical current and bias voltage, whose explicit analytic expressions are derived using the Lagrangian variational method. The dynamical phase diagram in nonlinear parameter space is presented, identifying oscillation regions with different skewed sinusoidal dependence, and diffusion regions with irreversible soliton spreading due to instability of the barrier. Furthermore, we obtain periodic dispersion relations of the solitons, indicating a switch between positive and negative inertial masses, consistent with the oscillation behaviors. Our results provide a general and comprehensive theoretical framework for soliton oscillation dynamics and pave the way for investigating various nonlinear transports and their potential applications.
\end{abstract}
\maketitle

\textcolor{blue}{\textit{Introduction.}}---The Josephson effect \cite{Josephson}, a quantum phenomenon describing supercurrents between two macroscopic systems (e.g., superconductors, superfluids or BECs) separated by a thin barrier, has been observed in various platforms \cite{Anderson,Backhaus,Wheatley,Leggett,Sukhatme,Hoskinson,Abbarchi,Levy,Dalfovo,Andrews,Smerzi,Ohberg,Williams,Raghavan,Cataliotti,Zibold,Kreula,Spagnolli,Burchianti,Burchianti2,Valtolina,Luick,Mukhopadhyay,Velkovsky}. A key feature of the Josephson effect is the periodic current (motion of Cooper pairs) under constant voltage (electric force), which has led to important device applications in quantum information processing \cite{Zhu,Kockum} and precise measurements \cite{Makhlin}. Remarkably, it was recently shown that the dynamics of a dark-bright soliton under constant forces can be mapped to the Josephson effect, where the bright soliton acts as a thin barrier separating the background superfluids of the dark soliton. The dark-bright soliton can oscillate periodically under a constant force, resulting in the periodic Josephson current across the effective barrier.

To date, studies on constant-force driven oscillations of solitons have mainly focused on nonlinear parameters with specific constraints \cite{Kosevich,Zhao,Yu,Bresolin}, such as the previously reported sinusoidal oscillations of the magnetic soliton in a two-component BEC \cite{Zhao} and the ferro-dark soliton in a spin-1 BEC \cite{Yu}. The interpretation of these soliton oscillations as Josephson effects is based on the mapping between spinor condensates and ferromagnetic systems characterized by the Landau-Lifshitz equation \cite{Bresolin}. This mapping, which gives rise to a perfect sinusoidal Josephson current (i.e., soliton oscillation), is valid only near the Manakov point with isotropic nonlinearity \cite{Manakov}. On the other hand, it has been shown that dark-bright solitons and other vector solitons can exist for more general nonlinear parameters \cite{Mao}. Therefore, a natural question arises: whether the constant-force driven soliton oscillations persist for general nonlinearities? Meanwhile, the Landau-Lifshitz equation breaks down far away from the Manakov point, though the mapping between the soliton oscillation (if exists) and the Josephson effect may still apply, the oscillation could significantly deviate from the sinusoidal form. A general Josephson equation characterizing such oscillations is still lacking.

In this letter, we address these questions by considering a two-component BEC system with general nonlinearities, and systematically investigate propagations of dark-bright solitons under constant forces. Our main results are: (i) By performing a Lagrangian variational method, we show that the soliton dynamics is fully captured by the self-adapted Josephson equation (SAJE), the analytical expressions of the self-adaptive critical current and bias voltage are presented explicitly. The self-adaption originates from the back-action of the Josephson current (i.e., soliton motion) on the barrier (i.e., bright soliton) that can freely evolve. (ii) Based on the Josephson equations, we analyze the stability of the effective barrier and find that the oscillation is a universal behavior that survives in a wide range of nonlinear parameter space, we also identify diffusion regions where the soliton spreads irreversibly without oscillation. We show that the nonlinear interactions enrich the soliton dynamics, giving rise to nonsinusoidal oscillations, which are classified into two types based on the skew direction. These results are confirmed by solving the Gross-Pitaevskii equation (GPE) directly. (iii) Finally, we calculate the dispersion relation of the soliton and observe the switch between positive and negative inertial masses, the asymmetric circles of dispersion relations correspond to nonsinusoidal oscillations.

\textcolor{blue}{\textit{The system.}}---We consider a two-component BEC with a long-cigar shape, the dynamics is described by quasi-one-dimensional coupled GPE
\begin{eqnarray}
	i \frac{\partial \psi_1}{\partial t} &=& -\frac{1}{2}\frac{\partial^2\psi_1}{\partial x^2} +(g_{11}|\psi_1|^2 + g_{12}|\psi_2|^2 + V_{1}) \psi_1, \label{eq:GP1}\\
	i \frac{\partial \psi_2}{\partial t} &=& -\frac{1}{2}\frac{\partial^2\psi_2}{\partial x^2} +(g_{12}|\psi_1|^2 + g_{22}|\psi_2|^2 + V_{2}) \psi_2 \label{eq:GP2}.
\end{eqnarray}
Here, $\psi_1$ and $\psi_2$ denote the wave functions of two condensate components, respectively. The above equations are dimensionless, the length, time and energy are in units of $\ell_0=\sqrt{\hbar/(m\omega_\bot)}$, $\tau_0=\omega_\bot^{-1}$, and $\epsilon_0=\hbar\omega_\bot$, where $\omega_\bot=\omega_y=\omega_z$ denotes the frequencies of the tight harmonic traps along the radial direction. The reduced quasi-one-dimensional nonlinear interaction strengths are $g_{{\rm ij}}=2a_{{\rm ij}}/\ell_0$ with $a_{{\rm ij}}$ representing the s-wave scattering lengths, which can be manipulated by the Feshbach resonances \cite{Feshbach1,Feshbach2,Feshbach3,Feshbach4,Feshbach5,Feshbach6}. We consider repulsive inter-component interaction $g_{12}>0$ and set $g_{12}=2$ throughout this paper (the physics is similar for different $g_{12}$). $V_{\rm i}$ represent the external spin-dependent potentials along $x$-direction.

Previous studies show that, the system supports the exact solution of a dark-bright soliton in the nonlinear region $(g_{11}-g_{12})(g_{22}-g_{12})<0$ \cite{Mao}, the solution reads $\psi_1=\sqrt{\frac{g_{12}-g_{22}}{g_{11}-g_{12}}}p \sech(\frac{x-x_c}{w}){\rm e}^{ivx}$, $\psi_2=i\sqrt{1-p^2} +p\tanh(\frac{x-x_c}{w})$ (we have assumed $\psi_1$ to be the bright soliton component without loss of generality), where $p^2$ is the dark-soliton notch depth, $x_c$, $w$ and $v$ are the soliton center, width and velocity, respectively. In the following, we will focus on the region $(g_{11}-g_{12})(g_{22}-g_{12})<0$ and investigate the propagation of these solitons under a constant weak force by applying the gradient potential $V_1=-F x$ and $V_2=0$ (the weak harmonic trap is omitted). {Note that the force should be weak to ensure that the potential is approximately constant over the soliton scale.}

\textcolor{blue}{\textit{Self-adapted Josephson equation.}}---We consider the initial states given by exact analytical soliton solution with $p(0)=1,v(0)=0,x_c(0)=0$, while $w(0)$ is uniquely determined by the nonlinearities $g_{{\rm ij}}$. The soliton starts to propagate in the presence of a constant force, and we assume its form maintains while the parameters [$p(t)$, $x_c(t)$, $w(t)$, etc.] become time-dependent. From the GPE, we can write down the Lagrangian of the system, the dynamical equation of the soliton parameters can be obtained by using the variational method (see details in the supplemental material \cite{SM}), which leads to two independent equations of motion
\begin{eqnarray}
	\dot{x}_c &=& \frac{\sqrt{1-p^2} (1-g_{12} N_B w +2g_{22}p^2w^2)}{3pw}, \label{vt}	\\
	F N_B &=& \frac{{\rm d}}{{\rm d}t}\left[ 2\arcsin (p) -2p\sqrt{1-p^2} +N_B\dot{x}_c \right], \label{motion}
\end{eqnarray}
the width of soliton satisfies $w(t) \!=\! \frac{g_{12}\! N_B +\sqrt{g_{12}^2 N_B^2 \!+\! 16g_{22} p^2}}{4g_{22} p^2}$, and $N_B$ is the particle number of the bright soliton.

Taking the bright soliton as an effective barrier that separating the background superfluids of the dark soliton \cite{Bresolin}, the motion of soliton would induce an effective current through the barrier. Since the range of oscillation motion is much larger than the soliton scale, the current can be written as $I(t) = \frac{{\rm d}}{{\rm d}t}\int_{x_c(t)}^{\infty}|\psi_2|^2 {\rm d}t = -n_0 \dot{x}_c(t)$, where $n_0=1$ is the normalized background density. According to the solution of the dark soliton, we find that the phase jump $\phi$ of the wave function $\psi_2$ across the barrier is related to the notch depth through $p=\sin(-\phi/2)$ with $\phi(0)=-\pi$. Substituting the above relations into the equation of motion given by Eqs.~\eqref{vt} and~\eqref{motion},  we finally arrive at the SAJE
\begin{eqnarray}
	I &=& \left[\frac{1}{N_B} -\frac{\lambda(\phi)}{N_B \sin(\phi)}\right] \sin(\phi) \equiv I_c(\phi) \sin(\phi) , \\ \label{current}
	\dot{\phi}&=& \frac{-F N_B}{1-\frac{{\rm d} \lambda(\phi)}{{\rm d} \phi}} \equiv U(\phi) N_B.
\end{eqnarray}
The implicit factor $\lambda(\phi)=N_B \dot{x}_c+\sin(\phi)$ is a function of $\phi$ and is responsible for the deviations from standard sinusoidal oscillation. The effective critical current $I_c(\phi)$ generally depends on the phase jump, since the profile of the bright soliton, which acts as a moving barrier \cite{Bresolin}, varies with its speed. Meanwhile, the effective bias voltage $U(\phi)$ also becomes $\phi$-dependent, although the external force is constant, since the accumulation of the phase difference originates from the change of soliton velocity, and the acceleration depends on the moving speed. In general, $\lambda(\phi)\neq 0$, and the oscillation is nonsinusoidal. Only when the interactions satisfy certain constraints, one may roughly have $\lambda(\phi)\simeq 0$, the oscillation reduces to nearly sinusoidal as studied in Refs.~\cite{Bresolin,Josephson,Zhao,Meng}.
\begin{figure}[t]
\begin{center}
\includegraphics[width=0.9\linewidth]{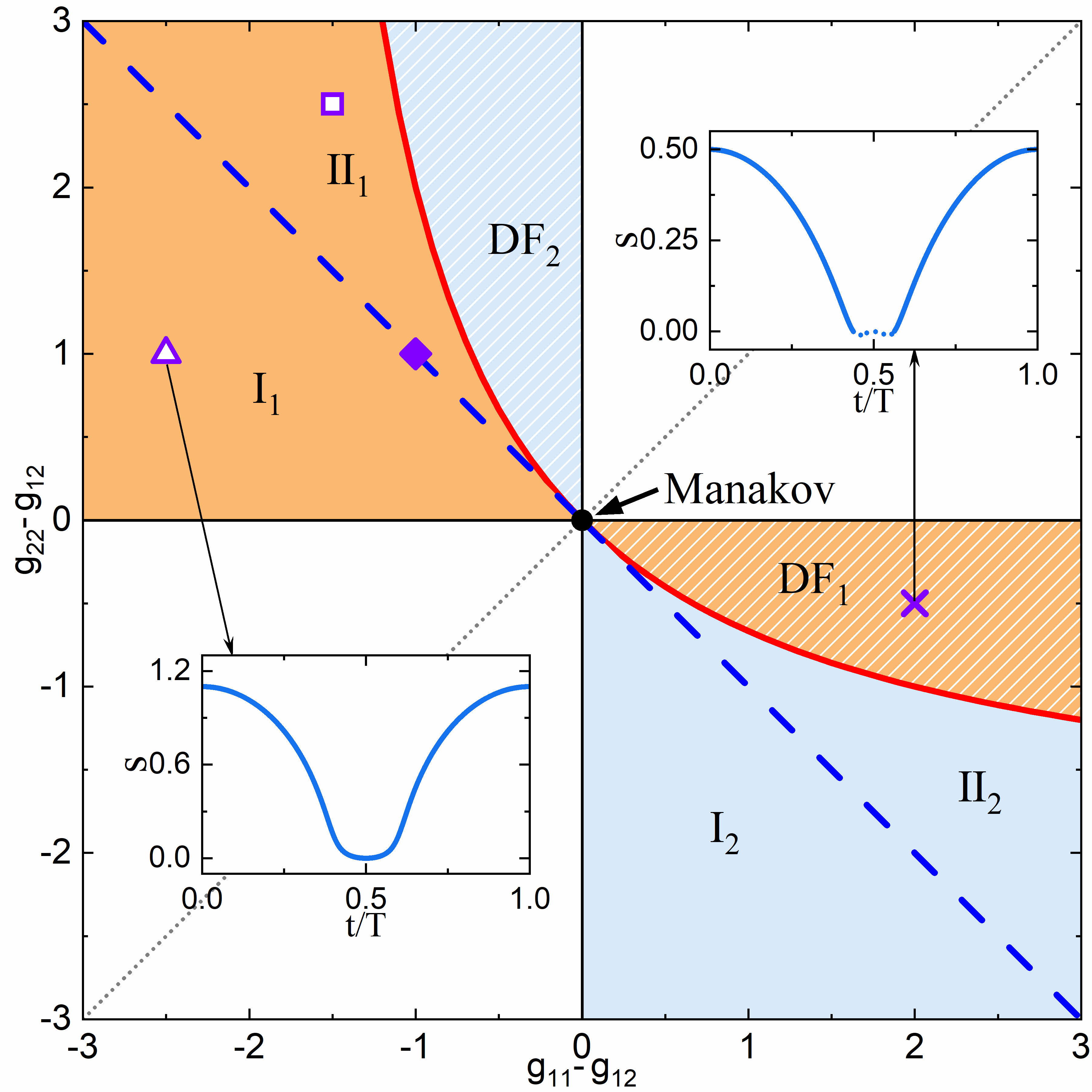}
\end{center}
\caption{Dynamical phase diagram for dark-bright solitons in the nonlinear parameter space. The red line separates the oscillation (clean, I$_{\rm i}$ and II$_{\rm i}$) and the diffusion (shaded, DF$_{\rm i}$) regions, and the blue dashed line separates two different oscillation phases (I$_{\rm i}$ and II$_{\rm i}$). The subscript ${\rm i}$ represents that, the ${\rm i}$-th component of the BEC is a bright soliton. The phase diagram is symmetric with respect to the thin dotted line ($g_{11}=g_{22}$) upon exchanging the two components. The black dot represents the Manakov case with isotropic nonlinearities. The insets show time evolutions of $s(t)$ for the triangle ($g_{11}=-0.5, g_{22}=3$) and the cross ($g_{11}=4, g_{22}=1.5$), where the dotted part indicates $s(t)<0$ and the barrier is unstable.} \label{fig:regions}
\end{figure}	

\textcolor{blue}{\textit{Dynamical phase diagram.}}---We distinguish different phases according to the oscillation and stability properties of the soliton dynamics. The SAJE always leads to periodic solutions for the Josephson current (i.e., oscillating soliton solutions), this is because we have assumed a stable barrier by requiring the soliton to maintain its form. However, due to the self-adaption, such requirement is no longer suitable if the barrier becomes unstable during propagation. To analyze the stability of the barrier, we examine the interaction-induced potential experienced by the bright soliton $V_\text{int}(x,t)=g_{11}|\psi_1|^2 + g_{12}|\psi_2|^2$. The barrier (i.e., the bright soliton) is stable if $V_\text{int}(x,t)$ always has a dip at $x_c$, and becomes unstable if the dip evolves into a hump at some time. We define the stable coefficient $s(t)=V_\text{int}(\pm\infty,t)-V_\text{int}(x_c,t)$ and instability is reflected by $s(t)<0$ during propagation.

In Fig.~\ref{fig:regions}, we plot the the dynamical phase diagram in the nonlinear parameter space spanned by $(g_{11}-g_{12})$-$(g_{22}-g_{12})$ plane. In the region $(g_{11}-g_{12})(g_{22}-g_{12})<0$ (colored regions) with exact static soliton solutions, we identify four oscillation phases (denoted by I$_1$, I$_2$, II$_1$, II$_2$) and two diffusion phases (denoted by DF$_1$, DF$_2$). The subscript ${\rm i}$ represents that, the ${\rm i}$-th component of the BEC is a bright soliton for the static solution without the force, and ${\rm i}=1$ (${\rm i}=2$) in the orange (blue) regions. The transition between diffusion phases and oscillation phases is obtained by analyzing the stability coefficient $s(t)$ [typical evolution of $s(t)$ in the diffusion and oscillation phases are shown in the insets], which leads to an analytic boundary given by the upper branch of $g_{12}^2=g_{11}g_{22}$ (thick red solid line in Fig.~\ref{fig:regions})~\cite{SM}. The oscillation survives in a wide range of nonlinear parameter space (lower left side of the phase boundary), it is a universal behavior not limited to the Manakov vicinity~\cite{Bresolin} or the special constraint condition $2g_{12}=g_{11}+g_{22}$ (thick blue dashed lines in Fig.~\ref{fig:regions})~\cite{Zhao}. The interaction also enriches the oscillations by self-adaption, the $I$-$t$ curve generally deviates from a simple sinusoidal dependence, and instead shows a skewed form. Based on the skew direction, we classify the oscillations into two types: I$_{\rm i}$ and II$_{\rm i}$. The positions of the maximum current are pulled towards $t=T/2 \text{ mod }T$ ($t=0\text{ mod }T$) for phase I$_{\rm i}$ (II$_{\rm i}$) as compared to the sine form, {with $T=|\frac{2\pi}{FN_B}|$ being the oscillation period \cite{Zhao,SM}.} The boundary between I$_{\rm i}$ and II$_{\rm i}$ is just the constraint condition $2g_{12}=g_{11}+g_{22}$ (thick dashed lines in Fig.~\ref{fig:regions}), at this boundary the peak current occurs at $t=\pm T/4$ mod $T$ and the $I$-$t$ curve is nearly sinusoidal. In contrast to the oscillation phase, the soliton spreads irreversibly in the diffusion phases DF$_{\rm i}$ (patterned regions in Fig.~\ref{fig:regions}), due to the instability of the barrier during propagation [i.e., $s(t)$ becomes negative at certain time]. In phase DF$_{\rm i}$, the intra-component repulsive interaction $g_{{\rm ii}}$ dominates, allowing the bright soliton $\psi_{\rm i}$ to diffuse easily. Note that both the diffusion and oscillation phases can exist in the vicinity of the integrable Makanov point, and surprisingly, the Makanov point locates on the phase boundary, which does not support soliton oscillation.
\begin{figure}[t]
\begin{center}
\includegraphics[width=\linewidth]{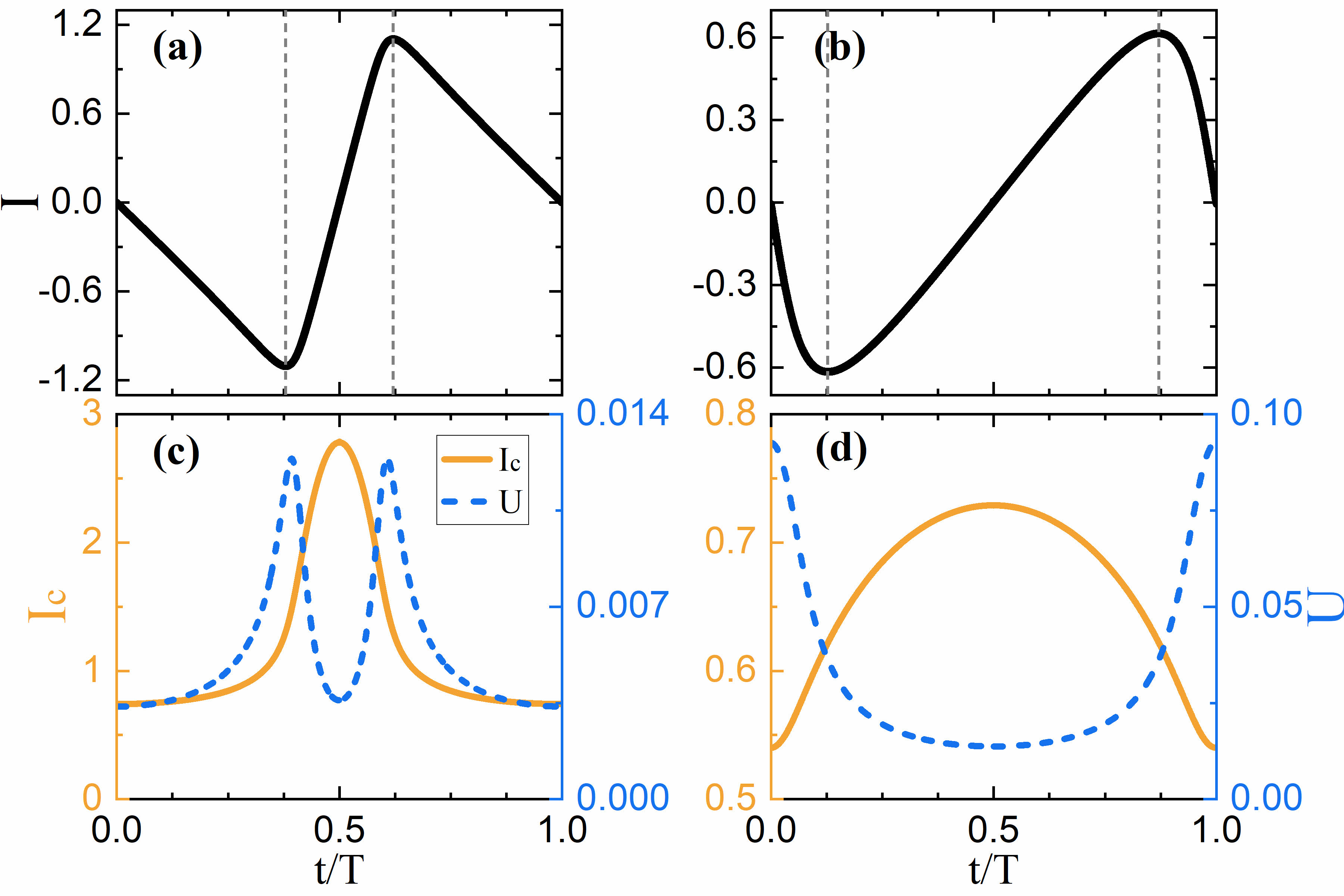}
\end{center}
\caption{The evolution of (a)-(b) Josephson current $I$ and (c)-(d) self-adapted critical current $I_c$ and bias voltage $U$ during one oscillation period. The parameters are given by the triangle ($g_{11}=-0.5, g_{22}=3$) and square ($g_{11}=0.5, g_{22}=4.5$) in Fig.~\ref{fig:regions}, respectively. The vertical dashed lines in (a) and (b) mark the maximum current which clearly demonstrate different skewed sinusoidal oscillations.} \label{fig:self}
\end{figure}

To see how interactions enrich the dynamics more directly, we plot the Josephson current $I$, self-adapted critical current $I_c$ and bias voltage $U$ as functions of $t$ from different oscillation phases in Fig.~\ref{fig:self}. It is worth noting that, when the two BEC components ${\rm i}=1$ and $2$ are swapped, the phase diagram is mirror symmetric with respect to $g_{11}=g_{22}$ (thin dotted line in Fig.~\ref{fig:regions}). Without loss of generality, we can only consider phases I$_1$ and II$_1$ where $\psi_1$ is the bright soliton. From Figs.~\ref{fig:self}(a) and \ref{fig:self}(b) we clearly see the different skewed sinusoidal oscillations under different interactions. Moreover, for both I$_1$ and II$_1$ phases, the critical current $I_c$ and bias voltage $U$ change dramatically within each period [as shown in Figs.~\ref{fig:self}(c) and \ref{fig:self}(d)], indicating the breakdown of the conventional Josephson equation description~\cite{Bresolin}.

\begin{figure}[t]
\begin{center}
\includegraphics[width=\linewidth]{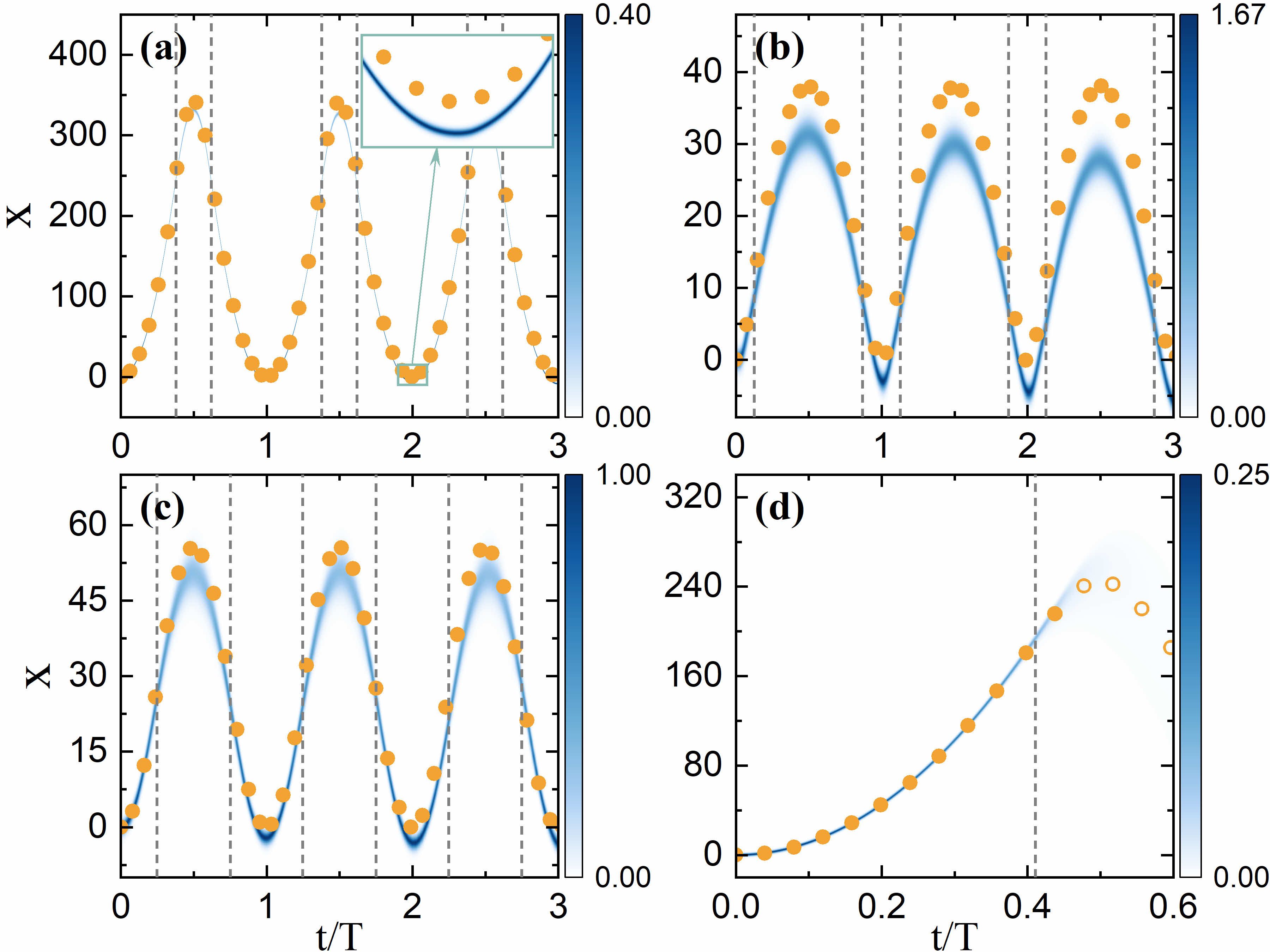}
\end{center}
\caption{Soliton dynamics obtained from GPE (density) and SAJE (orange circles) in different phases. Parameters in (a)-(d) correspond to the triangle, square, diamond ($g_{11}=1, g_{22}=3$) and cross in Fig.~\ref{fig:regions}, respectively. Filled (open) circles indicate that the values of $s(t)$ are positive (negative). In (a), the displacement is much larger than the soliton scale, the inset shows the zoom-in of the framed region. The vertical dashed lines mark the maximal and minimal soliton velocities to highlight the skewed oscillations.}\label{fig:oscillation}
\end{figure}

To confirm the predictions of our theory, we solve the GPE directly and compare them with those obtained from the SAJE. The GPE are numerically solved by using the integrating-factor method \cite{JYang} and imaginary-time propagation method \cite{Lehtovaara} with hard-wall traps. The results for different phases are plotted in Fig.~\ref{fig:oscillation}, where only bright solitons are shown, since the dark soliton follows the same trajectory as the bright one~\cite{SM}. The density evolutions obtained from the GPE are in good agreement with the trajectories obtained from the SAJE (circles in Fig.~\ref{fig:oscillation}), especially for phase I$_{\rm i}$ in Fig.~\ref{fig:oscillation}(a). For phase II$_{\rm i}$, slight deviations are observed due to the small particle loss and profile deformation of the soliton in realistic GPE simulation. Besides, the oscillation center slightly drifts towards the force direction in the GPE simulation, which is caused by the force-induced energy radiation and re-excitation of the dark soliton. For the DF$_{\rm i}$ phase shown in Fig.~\ref{fig:oscillation}(d), the GPE soliton solution follows the SAJE trajectory for the interval $s(t)>0$ (filled circles) and starts to irreversibly spread when $s(t)<0$ (open circles).

\begin{figure}[t]
\begin{center}
\includegraphics[width=\linewidth]{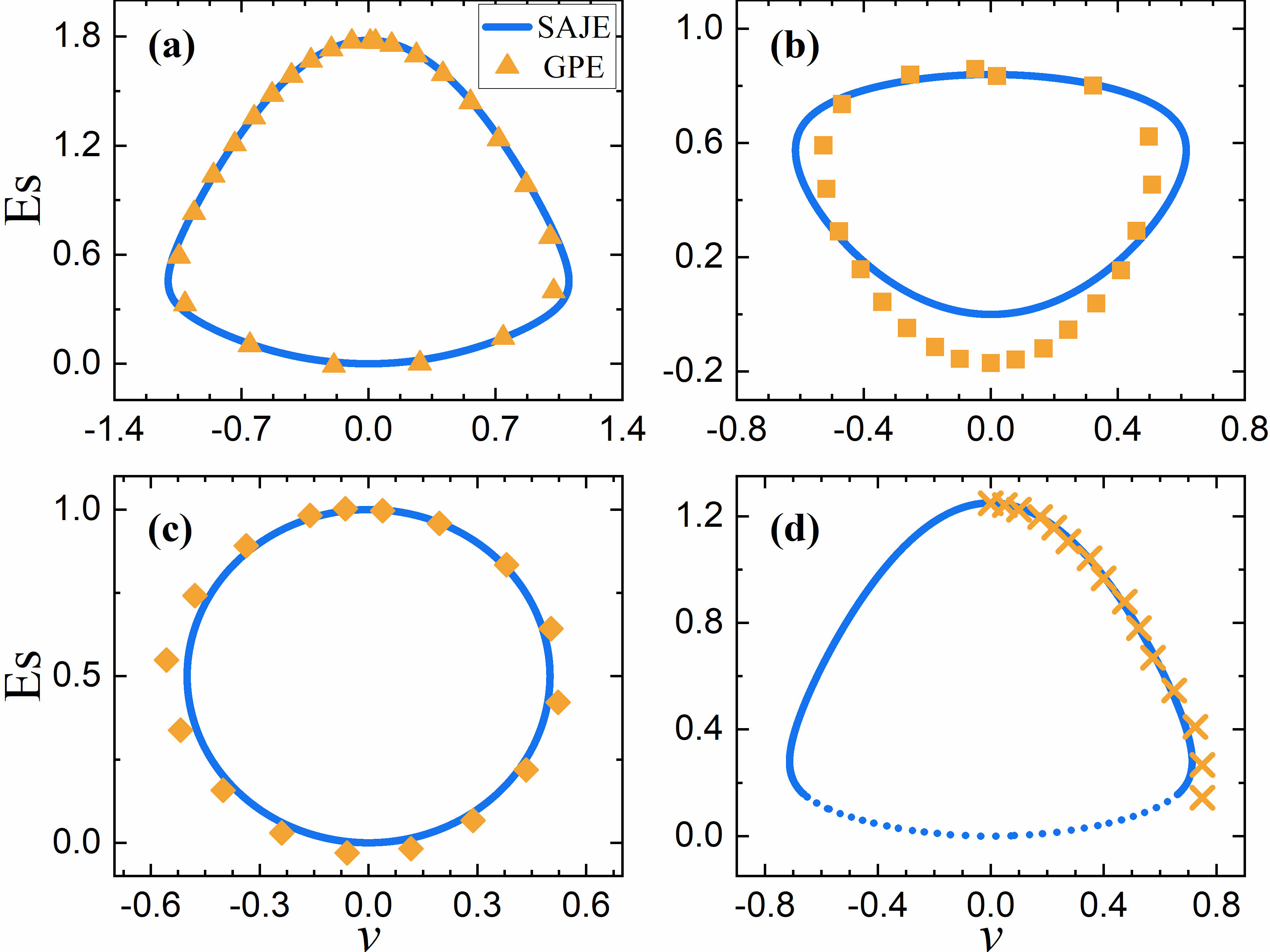}
\end{center}
\caption{(a)-(d) The dispersion relations for dark-bright solitons in different phases. The parameters in (a)-(d) are the same as that in Fig.~\ref{fig:oscillation}(a)-(d). The asymmetric circles in (a) and (b) directly imply the skewed oscillations and Josephson currents in Fig.~\ref{fig:self}. The dotted segment in (d) indicates $s(t)<0$ where the soiton becomes diffusive.}\label{fig:dis_rel}
\end{figure}

\textcolor{blue}{\textit{Dispersion relation and inertial mass.}}---The inertial mass of a soliton can be defined as $M^*=2\partial E_s/\partial(v^2)$ based on its dispersion relation \cite{Brazhnyi1,Scott,Brand}, with the soliton excitation energy $E_s = \int^{\infty}_{-\infty} [\frac{1}{2}|\partial_x \psi_1|^2 +\frac{1}{2}|\partial_x \psi_2|^2 +\frac{g_{11}}{2}|\psi_1|^4 +\frac{g_{22}}{2}(|\psi_2|^2-1)^2 +g_{12}|\psi_1|^2(|\psi_2|^2-1)]{\rm d}x$. Based on SAJE, the dispersion relation $E_s$-$v$ can be obtained, which is also periodic as shown in Fig.~\ref{fig:dis_rel} for different phases, the soliton oscillation corresponds to propagation along the periodic dispersion relation. The upper (lower) branch indicates negative (positive) inertial mass of the soliton, the two branches meet at the turning points with maximal and minimal soliton velocities (i.e., Josephson current). Therefore, the upper (lower) branch corresponds to the interval centered at $t=0$ ($t=T/2$) between the maximal and minimal Josephson currents. As a result, we see that in phase I$_{\rm i}$ [see Fig.~\ref{fig:dis_rel}(a)], the upper branch (i.e., negative mass) dominates over the lower branch (i.e., positive mass) in the dispersion relation, since the positions of the maximum current $I(t)$ are pulled toward $t=T/2$ compared to the sinusoidal form. A similar analysis applies to the II$_{\rm i}$ phase in Fig.~\ref{fig:dis_rel}(b). The dispersion relation at the I$_{\rm i}$-II$_{\rm i}$ boundary is also calculated, which forms an almost perfect circle as shown in Fig.~\ref{fig:dis_rel}(c). For the diffusion phase, the SAJE also gives rise to periodic dispersion relations, as shown in Fig.~\ref{fig:dis_rel}(d), including the stable (solid line) and unstable (dotted line) segments.

In Fig.~\ref{fig:dis_rel}, we also plot the dispersion relation obtained by numerically solving the GPE (see the markers), and the results are in good agreement with the SAJE results. The slight discrepancy in Fig.~\ref{fig:dis_rel}(b) is attributed to the profile deformation and particle loss of the soliton in real process characterized by GPE, the assumptions of $\sech$ and $\tanh$ soliton profile in deriving SAJE do not capture these features. In Fig.~\ref{fig:dis_rel}(d), the soliton cannot go through a full period, but the GPE result coincides with the SAJE result before the diffusion occurs.

\textcolor{blue}{\textit{Conclusion and discussion.}}---In summary, we have developed a general theoretical framework to study the dark-bright soliton propagation under constant forces, utilizing the mapping between soliton oscillation and Josephson effect. We derive the self-adapted Josephson equation analytically using the Lagrangian variational method, which can characterize the soliton dynamics very well. By analyzing the stability of the bright soliton (insulating barrier), we present the dynamical phase diagram in the nonlinear parameter space and identify different oscillation and diffusion phases. We show that, the soliton oscillations persist in a wide range of nonlinear interactions, which in turn enrich the oscillations and give rise to different skewed sinusoidal dependences. These interesting oscillations are also reflected in the dispersion relation and the inertial mass of the soliton.

In this work, we have mainly focused on the region $(g_{11}-g_{12})(g_{22}-g_{12})<0$ supporting exact soliton solutions. Although exact soliton solutions are not yet found in the region $(g_{11}-g_{12})(g_{22}-g_{12})>0$, stable solitons may exist, and thus, by assuming a proper variational ansatz, it is possible to generalize our results to arbitrary nonlinearities supporting stable solitons. {We expect that dark-bright solitons in the third quadrant exhibit stable oscillations when subjected to constant forces.
Conversely, those in the first quadrant are not anticipated to be stable, due to the intra-species interaction being larger than the inter-species interaction, the effective potential can evlolve into a hump.} Moreover, it would be interesting to consider the dynamics of multi-solitons in ring-shaped BECs, mimicking the self-adapted superconducting quantum interference device (SQUID) structure which may find potential applications in precise measurement. Therefore, our results may motivate further experimental and theoretical studies on various nonlinear transports and pave the way for exploring their applications in quantum devices.

\textcolor{blue}{\textit{Note added.}}---Unpon finishing our manuscript, we became aware of a recent experimental work~\cite{Rabec}, in which the oscillation dynamics were observed in the vicinity of the Manakov point, consisting with our theoretically predictions.

\section*{Acknowledgments}
We are grateful to Prof. Jie Liu for his helpful discussions. L.-C. Zhao is supported by the National Natural Science Foundation of China (Grant No. 12375005, 12235007). L.-Z. Meng is supported by the National Natural Science Foundation of China (Grant No. 12405002). X.-W. Luo is supported by Innovation Program for Quantum Science and Technology (Grant No. 2021ZD0301200), National Natural Science Foundation of China (Grant No. 12275203) and the USTC start-up funding.

\begin{widetext}
\section*{Supplemental Materials}
This supplemental material presents the Lagrangian variational method for deriving the self-adapted Josephson equation in explicit form, accompanied by a concise discussion of the oscillation period. An alternative explanation for the transition between the oscillation and diffusion phases is provided by analyzing the stability of the effective barrier (the bright soliton). Evolutions of densities for dark soliton components in a variety of scenarios are also presented.

\textcolor{blue}{\textit{Lagrangian variational method.}}---The oscillations of dark-bright solitons driven by constant forces have been reported in several researches investigating different Bose-Einstein condensate systems \cite{Kosevich,Zhao,Yu,Bresolin}. The dimensionless form for the coupled Gross-Pitaevskii equations governing the dynamics of a binary Bose-Einstein condensate can be written as
\begin{eqnarray}
i \frac{\partial \psi_1}{\partial t} &=& -\frac{1}{2}\frac{\partial^2\psi_1}{\partial x^2} +(g_{11}|\psi_1|^2 + g_{12}|\psi_2|^2 + V_{1}) \psi_1, \label{A1}\\
i \frac{\partial \psi_2}{\partial t} &=& -\frac{1}{2}\frac{\partial^2\psi_2}{\partial x^2} +(g_{12}|\psi_1|^2 + g_{22}|\psi_2|^2 + V_{2}) \psi_2 \label{A2}.
\end{eqnarray}
A recent study has revealed that the exact dark-bright soliton solution for this coupled model can exist in a much broader range of nonlinearities \cite{Mao}. Based on this progress, we have performed numerical simulations of dark-bright solitons with arbitrary nonlinear interaction strengths and observed universal oscillations under the effect of a constant force applied to the bright soliton component ($V_1=-F x$ and $V_2=0$). To gain insight into these oscillation behaviors, we employ the Lagrangian variational method to investigate the motion of dark-bright solitons. The trial functions can be assumed as follows
\begin{eqnarray}\label{trial_func1}
\psi_1 &=& f(t) \sech\left[\frac{x-x_c(t)}{w_1(t)}\right] \textrm{e}^{i \{\theta_0(t) + \theta_1(t)[x-x_c(t)]\}}, \\
\psi_2 &=& i\sqrt{1-p(t)^2} + p(t)\tanh\left[\frac{x-x_c(t)}{w_2(t)}\right],
\end{eqnarray}
where $f^2(t)$ and $p^2(t)$ denote the amplitude of the bright soliton and the notch depth of the dark soliton. $w_1(t)$ and $w_2(t)$ are the widths of the bright and dark solitons, respectively. In addition, the force should be weak to ensure that the potential is approximately constant over the soliton scale, and thus the potential energy can be reduced as $E_p=\int_{-\infty}^{+\infty}Fx|\psi_1|^2{\rm d}x = F x_c \int_{-\infty}^{+\infty}|\psi_1|^2{\rm d}x$, where $x_c$ is the soliton center. The Lagrangian for dynamical equations Eqs.~\eqref{A1} and \eqref{A2} can be expressed as \cite{Kivshar}
\begin{eqnarray}\label{L1}
\mathcal{L}(t) &=& \int_{-\infty}^{+\infty} \bigg[\frac{i}{2}(\psi_1^* \partial_t \psi_1 -\psi_1 \partial_t \psi_1^*) +\frac{i}{2}(\psi_2^* \partial_t \psi_2 -\psi_2 \partial_t \psi_2^*)(1-\frac{1}{|\psi_2|^2}) -\frac{1}{2}|\partial_x \psi_1|^2 - \frac{1}{2}|\partial_x \psi_2|^2 \nonumber \\
&& -\frac{g_{11}}{2}|\psi_1|^4 -\frac{g_{22}}{2}(|\psi_2|^2-1)^2 -g_{12}|\psi_1|^2(|\psi_2|^2-1) +Fx|\psi_1|^2 \bigg] \textrm{d}x \nonumber \\
&=& 2f^2(t)w_1(t)[\theta_1(t)\dot{x}_c(t) -\dot{\theta}_0(t)] -2p(t)\sqrt{1-p^2(t)}~\dot{x}_c(t) +2\arcsin[p(t)]\dot{x}_c(t) -\frac{f^2(t)}{3w_1(t)} -f^2(t)w_1(t)\theta_1^2(t) \nonumber\\
&& -\frac{2p^2(t)}{3w_2(t)} -\frac{2g_{11}}{3}f^4(t)w_1(t) -\frac{2g_{22}}{3}p^4(t)w_2(t) +g_{12}f^2(t)p^2(t)\Gamma(w_1,w_2) +2Ff^2(t)w_1(t)x_c(t),
\end{eqnarray}
where $\Gamma(w_1,w_2)=\int_{-\infty}^{\infty} \{\sech^2[\frac{x-x_c(t)}{w_1(t)}]\sech^2[\frac{x-x_c(t)}{w_2(t)}]\}{\rm d}x$.
The soliton velocity is $v = \dot{x}_c(t) = \frac{{\rm d}x_c(t)}{{\rm d}t}$.
Subsequently, the application of the Euler-Lagrange equation $\frac{{\rm d}}{{\rm d}t}[\frac{\partial \mathcal{L}(t)}{\partial\dot{\alpha}}] = \frac{\partial \mathcal{L}(t)}{\partial\alpha}$, where $\alpha = f(t), p(t), w_1(t), w_2(t), x_c(t), \theta_0(t), \theta_1(t)$, yields
\begin{eqnarray}
	&2g_{12}f(t)p^2(t)\Gamma(\!w_1\!,\!w_2\!) \!-\!\frac{2f\!(t)}{3w_1\!(t)} \!-\!\frac{8g_{11} f^3\!(t)w_1\!(t)}{3} \!-\!2f\!(t)w_1\!(t)\theta_1^2\!(t) \!+\!4f\!(t)w_1\!(t)[\theta_1\!(t)\dot{x}_c\!(t)\!-\!\dot{\theta}_0\!(t)] \!+\!4F\!f\!(t)w_1\!(t)x_c\!(t)= 0, \\
	&2g_{12}f^2(t)p(t)\Gamma(w_1,w_2) -\frac{4p(t)}{3w_2(t)} -\frac{8g_{22} p^3(t)w_2(t)}{3} +\frac{2\dot{x}_c(t)}{\sqrt{1-p^2(t)}} +\frac{2p^2(t)\dot{x}_c(t)}{\sqrt{1-p^2(t)}} -2\sqrt{1-p^2(t)}~\dot{x}_c(t) = 0, \label{for2} \\
	&g_{12}f^2(t)p^2(t)\frac{\partial\Gamma(w_1,w_2)}{\partial w_1(t)} +\frac{f^2(t)}{3w_1^2(t)} -\frac{2g_{11} f^4(t)}{3} -f^2(t)\theta_1^2(t) +2f^2(t)[\theta_1(t)\dot{x}_c(t)-\dot{\theta}_0(t)] +2Ff^2(t)x_c(t) = 0, \\
	&g_{12}f^2(t)p^2(t)\frac{\partial\Gamma(w_1,w_2)}{\partial w_2(t)} -\frac{2g_{22} p^4(t)}{3} +\frac{2p^2(t)}{3w_2^2(t)} = 0, \label{for3} \\
	&\frac{{\rm d}}{{\rm d}t}[2\arcsin p(t) -2p(t)\sqrt{1-p^2(t)} +2f^2(t) w_1(t)\theta_1(t)] = 2Ff^2(t) w_1(t), \label{for4} \\
	&2f^2(t) w_1(t) \equiv {\rm constant}, \label{for1} \\
	&\theta_1(t) = \dot{x}_c(t).
\end{eqnarray}
Upon setting $w_1(t)=w_2(t)=w(t)$, we can obtain the reduced form of the integrals
\begin{eqnarray}
\Gamma &=& \int_{-\infty}^{\infty} \left\{\sech^4\left[\frac{x-x_c(t)}{w(t)}\right]\right\}{\rm d}x = \frac{4}{3}w(t), \\
\frac{\partial\Gamma}{\partial w_1(t)} = \frac{\partial\Gamma}{\partial w_2(t)} &=& \int_{-\infty}^{\infty} \left\{ \frac{2[x-x_c(t)]}{w^2(t)} \tanh\left[\frac{x-x_c(t)}{w(t)}\right]\sech^4\left[\frac{x-x_c(t)}{w(t)}\right]\right\} {\rm d}x = \frac{2}{3}.
\end{eqnarray}
Finally, three fundamental functions for describing the motion of the soliton are presented as follows
\begin{eqnarray}
	&&\dot{x}_c(t) = \frac{\sqrt{1-p^2(t)} [1-g_{12} N_Bw(t) +2g_{22}p^2(t)w^2(t)]}{3p(t)w(t)} = \frac{4g_{22}p(t)\sqrt{1-p^2(t)}}{g_{12}N_B +\sqrt{g_{12}^2N_B^2 +16g_{22}p^2(t)}}, \label{Ev-v}
	\\
	&&w(t) = \sqrt{\frac{1}{g_{22}p^2(t) -g_{12}f^2(t)}} = \frac{g_{12}N_B +\sqrt{g_{12}^2N_B^2 +16g_{22}p^2(t)}}{4g_{22}p^2(t)}, \label{Ev-w}\\
	&&2\arcsin [p(t)] -2p(t)\sqrt{1-p^2(t)} +N_B\dot{x}_c(t) = F N_B t +C, \label{motion_SM}
\end{eqnarray}
where $N_B \equiv \int_{-\infty}^{\infty}|\psi_1|^2{\rm d}x=2f^2(t) w(t)$ is the particle number of the bright soliton. The integration constant $C$ can be determined based on the initial state. It can therefore be stated that the motion of dark-bright solitons can be described theoretically, by solving the above three functions. It should be noted that the right-hand side of Eq.~\eqref{motion_SM} is related to the external potential $V_1$, and the initial states can be given by the exact analytical solution \cite{Mao}
\begin{eqnarray}
&&\psi_1\mid_{t=0}=f_0 \sech\left(\frac{x}{w_0}\right){\rm e}^{i v_0x}, \\
&&\psi_2\mid_{t=0}=i\sqrt{1-p_0^2} +p_0\tanh\left(\frac{x}{w_0}\right),
\end{eqnarray}
where $f_0=f(0)=\sqrt{\frac{g_{12}-g_{22}}{g_{11}-g_{12}}}p_0$, $w_0=w(0)=\sqrt{\frac{g_{12}-g_{11}}{g_{12}^2-g_{11}g_{22}}}\frac{1}{p_0}$ and $v_0=\dot{x}_c(0)=\pm\sqrt{\frac{g_{12}^2-g_{11}g_{22}}{g_{12}-g_{11}}(1-p_0^2)}$. The particle number of the bright soliton is $N_B = 2f_0^2w_0 = \sqrt{\frac{4(g_{12} -g_{22})^2}{(g_{12} -g_{11})(g_{12}^2 -g_{11}g_{22})}p_0^2}$. In the case of the initially static soliton, the initial depth of the dark soliton is $p_0=p(0)=1$ and resulting in $C=\pi$.

Upon setting $p(t) = \sin[-\phi(t)/2]$, the phase difference over the dark soliton can be expressed as
\begin{eqnarray}\label{delta_theta}
\phi(t) = \phi_{DS}(+\infty) - \phi_{DS}(-\infty) = -[F N_B t - \lambda(t)+C],
\end{eqnarray}
where $\lambda(t)=N_B \dot{x}_c(t) +\sin\left[\phi(t)\right]$ is introduced for the purpose of highlighting the deviations from the standard sinusoidal oscillation. By the way, it can be demonstrated that the relation between the velocity and width is $\dot{x}_c(t) w(t) = -\cot[\phi(t)/2]$, which is analogous to the relation found for the integrable Manakov case \cite{Busch}.

In order to gain insight into the nature of the oscillation dynamics of dark-bright solitons, we can then focus on the current across the moving barrier (the bright soliton). Since the soliton moves in a wide range, the influence of the local soliton structure can be ignored safely. The particle current is then directly calculated from its motion as \cite{Bresolin} $I(t) = \frac{{\rm d}}{{\rm d}t}\int_{x_c(t)}^{\infty}|\psi_2|^2 {\rm d}t = -n_0 \dot{x}_c(t)$, where $n_0=1$ is the normalized background density. Based on all above analyses, we can derive the effective Josephson equations,
\begin{eqnarray}
	I &=& I_c(\phi) \sin(\phi), \\
	\dot{\phi} &=& U(\phi) N_B,
\end{eqnarray}
where the effective critical current $I_c = [\frac{1}{N_B} -\frac{\lambda(\phi)}{N_B \sin(\phi)}] \sin(\phi) = -\frac{2g_{22}}{\sqrt{g_{12}^2N_B^2 +16g_{22}\sin^2(\phi/2)}}$, the effective bias voltage $U(\phi) = \frac{-F N_B}{1-\frac{{\rm d} \lambda(\phi)}{{\rm d} \phi}}$, and $\lambda(\phi) = N_B \dot{x}_c(\phi) +\sin(\phi)$. The period $T=|\frac{2\pi}{FN_B}|$ has been proposed for spin solitons on the thick blue dashed line in Fig.~\ref{fig:regions} of the main text \cite{Zhao}. It is imperative to underscore that this oscillation period is applicable to any dark-bright soliton beyond this limit, even if the current $I$ is skewed, since $I_c$ also varies with a period of $T$.

\textcolor{blue}{\textit{Transition between the oscillation and diffusion phases.}}---The preceding analyses indicate that the oscillation of dark-bright solitons exists for arbitrary nonlinear parameters. However, numerical simulations demonstrate the diffusion dynamics in DF$_{{\rm i}}$ regions. Here we provide an alternative way to understand this irreversible diffusion. The bright soliton acts as a moving barrier for the self-adapted Josephson effect. It is evident that the instability of the barrier during propagation leads to the invalidation of the periodicity of the Josephson current. As known to all, a bright soliton can exist stably due to the balance between the kinetic energy and the effective trap formed by the nonlinear term. In the event that this effective potential evolves into a localized hump during the propagation process, the barrier (bright soliton) will be unstable, thereby inducing irreversible soliton spreading. Therefore, we can study this interaction-induced potential experienced by the bright soliton, which can be expressed as $V_{{\rm int}} = g_{11}|\psi_1|^2+g_{12}|\psi_2|^2 = g_{12} +[\frac{g_{11}N_B}{2w} -g_{12}\sin^2(\frac{\phi}{2})]\sech^2(\frac{x}{w})$. Based on this, a coefficient can be defined to describe the structure of the potential
\begin{eqnarray}\label{st}
s(t) = V_{{\rm int}}(\pm\infty,t) - V_{{\rm int}}(x_c,t) = g_{12}\sin^2\!\!\left(\frac{\phi}{2}\right) -\frac{g_{11}N_B}{2w}.
\end{eqnarray}
The potential is a dip (hump) if $s(t)>0$ [$s(t)<0$] and then the barrier is stable (unstable). We note that for the initial states given by the exact dark-bright soliton, the effective potential $V_{{\rm int}}(t=0)$ is always a dip. However, if a real value of $\phi(t)$ exists to render $s(t)<0$, the interaction-induced potential experienced by the bright soliton can evolve into a hump and cause the soliton to diffuse. Conversely, if no real root $\phi(t)$ exists for $s(t)<0$, the effective potential will not become a hump. It is evident that $s(t)$ will always take a positive value if $g_{11}<0$, signifying that the intra-species interaction of the $\psi_1$ component is attractive. If $g_{11}>0$, the inequality $s(t)\leqslant0$ can be solved by substituting Eq.~\eqref{Ev-w} into Eq.~\eqref{st}, thereby deriving the condition for $\phi(t)$ to have real roots as
\begin{eqnarray}
g_{12}^2<g_{11}g_{22},
\end{eqnarray}
the upper branch of which corresponds to of the phase boundary (thick red solid line in Fig.~\ref{fig:regions} of the main text).
\begin{figure}[h]
\begin{center}
\includegraphics[width=\linewidth]{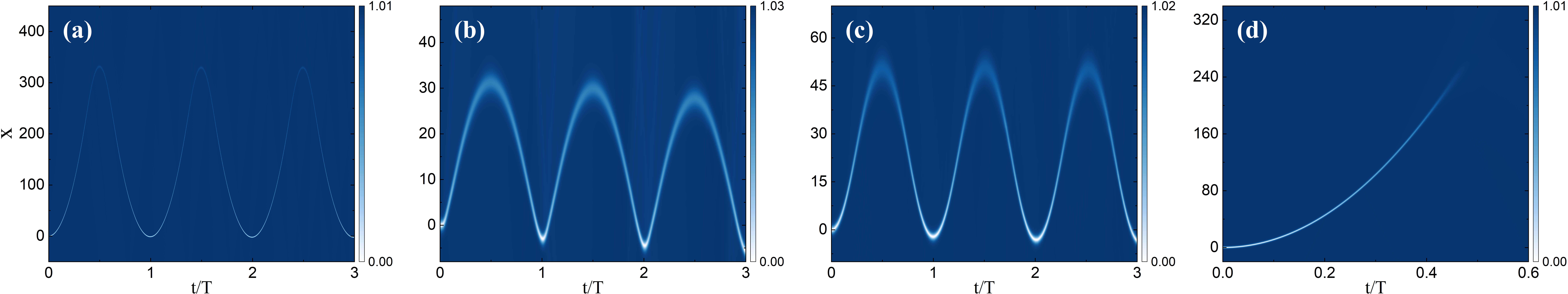}
\end{center}
\caption{Numerical time evolutions for the dark solitons corresponding to the bright solitons in Fig.~\ref{fig:oscillation} of the main text. The parameters are (a) $g_{11}=-0.5$, $g_{22}=3$, (b) $g_{11}=0.5$, $g_{2}=4.5$, (c) $g_{11}=1$, $g_{22}=3$, and (d) $g_{11}=4$, $g_{22}=1.5$. The inter-species interaction is $g_{12}=2$.}\label{fig:evolution}
\end{figure}

\textcolor{blue}{\textit{Density evolutions of dark soliton componments.}}---In Fig.~\ref{fig:oscillation} of the main text only the density evolutions of the bright soliton components are shown. Here, we clearly show the evolutions of the corresponding dark solitons in Fig.~\ref{fig:evolution}, when the linear potential $V_1=-F x$ is applied to the bright soliton. Dark solitons have exactly the same trajectories as the bright solitons. The numerical simulations have been conducted utilizing the integrating-factor method \cite{JYang} and the imaginary-time propagation method \cite{Lehtovaara}. The recent advancements in experimental technology have enabled the creation of a flat-top condensate with a nearly uniform density \cite{Meyrath,Gaunt,Chomaz,Navon,Roccuzzo,Ren}.
Consequently, it is reasonable to utilize a hard-wall trap $V_{HW}=100[\tanh(x-x_{b1}) - \tanh(x-x_{b2})+2]$, where $x_{b1}$ and $x_{b2}$ represent trap boundaries. Nonlinear interactions are considered in different regions, as illustrated in Fig.~\ref{fig:regions} of the main text (i.e., the triangle, square, diamond, and cross), with the force set to $F=-0.01$.
\end{widetext}

\end{document}